\documentclass{aa}  

\usepackage{graphicx}
\usepackage{txfonts}
\usepackage{lipsum}
\usepackage{subcaption}         % necessary for continued figures, example in section 3
\usepackage{lscape}             % to rotate a single page table, example in appendix.
\usepackage{placeins}           % useful with \FloatBarrier, to keep 
\usepackage[]{hyperref}
\usepackage{amsmath,amssymb,bm}
\usepackage{physics}
\usepackage{empheq}
\usepackage{xcolor}

\begin{document}

%%%%%%%%%%%%%%%%%%%%%%%%%%%%%%%%%%%%%%%%
% if you use custom commands in your title,
% ensure to check your title when submitting!
%%%%%%%%%%%%%%%%%%%%%%%%%%%%%%%%%%%%%%%%
\title{Analytical modeling of the gamma-ray morphology of TeV halos around moving pulsars}

\subtitle{}

%%%%%%%%%%%%%%%%%%%%%%%%%%%%%%%%%%%%%%%%
% Please separate each author with the \and command
%
% Use the \corrauth to provide the corresponding
% author address. It will be automatically inserted as 
% footnote in the PDF output.
%
% Please DO NOT include ORCIDs next to author names.
% Instead, please provide an active address for each coauthor:
% it will be automatically extracted by EDPS editorial system, 
% and co-authors will be be able to authenticate their ORCID.
%
% Only authenticated ORCIDs will be taken into account.
% ORCIDs included here will be removed.
%%%%%%%%%%%%%%%%%%%%%%%%%%%%%%%%%%%%%%%%

   \author{S. Recchia\inst{1}\email{sarah.recchia@ifj.edu.pl}        % use \corrauth for the corresponding author
   %\and A. A\inst{1}\email{}\fnmsep\thanks{}    
}

   \institute{Institute of Nuclear Physics Polish Academy of Sciences, PL-31342 Krakow, Poland 
  % \and 
   }

   \date{Received September 30, 20XX}

% \abstract{}{}{}{}{}
% 5 {} token are mandatory
 
\abstract
% context heading (optional)
{The discovery of extended TeV gamma-ray halos around a few middle-aged pulsars has provided new insights into the propagation of cosmic-ray leptons in the source surroundings. Previous studies have shown that pulsar proper motion can significantly
affect the morphology of TeV halos, producing asymmetric gamma-ray emission extending behind the pulsar trajectory. Such morphological signatures are potentially accessible to current and future gamma-ray observatories.}
% aims heading (mandatory)
{We aim to derive an analytical prescription for the line-of-sight-integrated
particle distribution around moving pulsars, suitable for constructing
spatial templates of the associated TeV-halo gamma-ray emission
without resorting to computationally expensive numerical integrations.}
% methods heading (mandatory)
{We consider electrons and positrons continuously injected by a moving source, undergoing isotropic diffusion and radiative energy losses.
Assuming spatially uniform target photon fields, as appropriate for the cosmic microwave background and approximately valid for the interstellar radiation fields, the projected particle distribution
provides a useful proxy for the inverse-Compton gamma-ray morphology.}
% results heading (mandatory)
{We show that the line-of-sight integration of the particle distribution can be performed exactly, reducing the calculation to a single numerical
integral over the injection history. Under the additional assumptions of constant injection luminosity and a simplified treatment of energy
losses, we derive a fully analytical expression that factorizes into an energy-dependent spectral term and a dimensionless spatial term
describing the energy-dependent morphology.
Comparison with the exact calculation shows good agreement over a broad range of energies and spatial scales, with appreciable deviations primarily at particle energies approaching the cutoff of the injection spectrum and at large distances from the source, where the emission
becomes fainter.}
% conclusions heading (optional)
{Our results enable the  construction of physically motivated, energy-dependent spatial templates for gamma-ray halos around moving
pulsars, capturing the morphological asymmetries induced by proper
motion without computationally expensive numerical integrations. This makes them particularly suitable for systematic morphological
and spectral analyses of extended gamma-ray sources.}
{}
  

   \maketitle
   \nolinenumbers

%%%%%%%%%%%%%%%%%%%%%%%%%%%%%%%%%%%%%%%%%%%%%%%%%%%%%%%%%%%%%%
\section{Introduction}
The first TeV halos were discovered in 2017 by the High-Altitude Water Cherenkov Observatory (HAWC), which reported \citep{HAWC-2017-Geminga-Monogem} extended very-high-energy (VHE, $>0.1$ TeV) gamma-ray emission, at photon energies of approximately 8--40 TeV, around the nearby middle-aged pulsars Geminga and PSR B0656+14 (Monogem). Both sources are located at distances of a few hundred parsecs from Earth, and their gamma-ray emission extends over approximately 20--30 pc. Subsequent observations by the Large High Altitude Air Shower Observatory (LHAASO) have revealed additional extended gamma-ray sources associated with other middle-aged pulsars \citep{LHAASO-2021-TeV-halo-PSR-J0622, LHAASO-2025-TeV-halo-PSRJ0248+6021}, and a number of analyzes have been devoted to search for TeV-halo-like structures around this class of astrophysical accelerators \citep{DiMauro-2020-low-diff-bubbles-PSR, DiMauro-2021-FermiLAT-HAWC-bright}. The natural interpretation of this emission is inverse-Compton scattering (ICS) by electrons and positrons released by the pulsars and propagating in the surrounding interstellar medium (ISM). At multi-TeV electron energies, the Klein--Nishina suppression makes the cosmic microwave background (CMB) the dominant target radiation field, unless a sufficiently intense local infrared field is present \citep[see e.g.][]{Breuhaus22}, while at lower energies, other components of the interstellar radiation field (ISRF) may also contribute \citep{Fang-2021-Klein-Nishina}. 
Since the CMB and diffuse ISRF provide uniform target photon fields, the gamma-ray morphology at a given photon energy largely traces the spatial distribution of the leptons within the characteristic energy range that
mostly  contributes to the corresponding inverse-Compton emission.

One of the most striking results of the HAWC observations was the inference of a diffusion coefficient approximately two to three orders of magnitude smaller than the average Galactic value inferred from cosmic-ray (CR) measurements \cite{Gabici-2019-review}, most notably the boron-to-carbon ratio, and extrapolated to multi-TeV energies. Indeed, within the interpretation of the emission in terms of isotropic, loss-limited diffusion, the observed extension of TeV halos requires $D(100\,{\rm TeV})\sim10^{27}$--$10^{28}\,{\rm cm^2\,s^{-1}}$. 
The possible presence of regions of suppressed diffusion around CR
accelerators has important implications for particle transport and
may influence the locally measured CR fluxes
\cite{Jacobs-Mertsch-2023-lowdiff-regions-Be10}.
Such regions may also affect the diffuse gamma-ray emission and
the contribution of pulsars to the CR positron excess
\citep[see][for reviews]{Liu-2022-TeV-halos-review, Amato-Recchia-2024-TeV-halos-review}.

Several explanations have been proposed for these findings. Some studies have focused on explaining suppressed diffusion in terms of the properties of the pulsar environment, which may be more turbulent than the Galactic average \citep{Lopez-Coto-Giacinti-2018-TeV-halos-coherence, Fang-2019-TeV-halos-slow-diff-origin}, or in terms of the amplification of magnetic turbulence by streaming instabilities driven by the escaping electron--positron pairs \citep{Evoli-2018-TeV-halos-selfgen,Mukhopadhyay-Linden-2022-TeV-halos-selfgen}. Other studies have questioned the assumption of isotropic suppressed diffusion, considering, for instance, the possible impact on gamma-ray emission of the transition from quasi-ballistic to diffusive propagation inherent to any diffusion scenario \citep{Recchia-2021-TeV-halos-ballistic,Bao-2022-TeV-halos-ballistic}, or, more importantly, the effect of anisotropic diffusion with respect to the large-scale magnetic field in the source region \citep{Lopez-Coto-Giacinti-2018-TeV-halos-coherence, Liu-2019-TeV-halos-anisotropic, DeLaTorre-2022-TeV-halos-anisotropic}. In this scenario the parallel diffusion coefficient can remain close to typical Galactic values, while diffusion perpendicular to the local field is strongly suppressed. This possibility has recently received renewed attention in light of new analyses of the Geminga and Monogem halos
\citep{Wu-2026-Geminga-anisotropic-morphology}.

The implications of suppressed diffusion around pulsars for the
CR positron excess have also been extensively investigated, often adopting two-zone diffusion models in which a region of inhibited diffusion surrounds the source, while particle propagation at larger distances is governed by the typical Galactic diffusion coefficient \citep{Profumo-2018-TeV-halos-two-zone-posfrac, Johannesson-2019-posfrac-twozone, Tang-Piran-2019-posfrac-Geminga, DiMauro-2019-FermiLAT-Geminga, Martin-2022-TeV-halos-rare-posfrac, Schroer-2023-TeV-halos-posfrac}.

An additional ingredient that may substantially affect the TeV-halo morphology is the pulsar proper motion. Due to natal kick velocities that can exceed several hundred km/s \citep{Igoshev-2020-pulsar-kick-velocity}, pulsars can travel appreciable distances during their lifetimes, but also during the time over which the emitting particles lose their energy through synchrotron and inverse-Compton losses. Consequently, the observed emission results from the superposition of particle populations injected at different positions along the pulsar trajectory \citep{Tang-Piran-2019-posfrac-Geminga}, potentially producing asymmetric halos with emission extending
behind the moving pulsar along its trajectory. This effect has already been investigated in the case of Geminga, including analyses of its lower-energy gamma-ray emission with Fermi-LAT \citep{DiMauro-2019-FermiLAT-Geminga}, and in subsequent studies of the morphology of gamma-ray halos around moving pulsars \citep{Zhang-2021-TeV-halo-moving-pulsar}.

The possible relevance of the pulsar proper motion can be appreciated by comparing the characteristic diffusion distance of the particles with the distance traveled by the pulsar during their radiative loss time. For particles of energy $E$, these scales are approximately
\begin{align*}
   & r_d(E)\simeq 2\sqrt{D(E)\tau_{\rm loss}(E)},\\
   & r_{\rm psr}(E)\simeq v_{\rm psr}\tau_{\rm loss}(E),
\end{align*}
where $D(E)$ is the diffusion coefficient, $v_{\rm psr}$ is the pulsar velocity, and $\tau_{\rm loss}(E)$ is the energy-loss time. For particles with energies $E \approx 10$--$200$ TeV, the latter is dominated by ICS losses on the CMB and synchrotron losses in the background magnetic field, giving a  loss time \citep{Atoyan-1995-long}:
\begin{equation*}
    \tau_{\rm loss} (E) \approx 3\times 10^4\, \left(\frac{10\, \rm TeV}{E}\right) \frac{1}{0.26 + 0.22 \left(\frac{B}{3\, \rm uG}\right)^2}\, \rm yr.
\end{equation*}
Assuming a Geminga-like diffusion coefficient, $D(100\,{\rm TeV})\sim 5\times10^{27}\,{\rm cm^2\,s^{-1}}$, electrons with energies $E\sim100$ TeV would diffuse over a distance of $\sim 20 \,\rm pc$, while a pulsar moving at $v_{\rm psr}\sim 100\,{\rm km\,s^{-1}}$ would travel about 1 pc. Larger pulsar velocities, which are not uncommon, can appreciably increase this displacement, as in the case analysed in \cite{LHAASO-2025-TeV-halo-PSRJ0248+6021}, where a proper-motion velocity of $\sim 500 \, \rm km/s$ was reported. On the other hand, for a diffusion coefficient that increases with energy as $D(E)\propto E^\delta$, with typical values $0.3 \lesssim\delta \lesssim 0.7$ \citep[see e.g.][]{Gabici-2019-review}, the ratio $r_{\rm psr}/r_d$ tends to increase at lower energies. A similar trend is expected for particles whose loss time exceeds the pulsar age, in which case $\tau_{\rm loss}(E)$ is replaced by $t_{\rm age}$.
Thus, depending on the pulsar velocity and the diffusion coefficient,
proper motion can significantly affect the halo morphology, producing
asymmetries that could be detectable in gamma-ray observations
\cite{Zhang-2021-TeV-halo-moving-pulsar}.

Although the morphology of gamma-ray halos around moving pulsars has already been investigated \cite{DiMauro-2019-FermiLAT-Geminga, Zhang-2021-TeV-halo-moving-pulsar}, the inclusion of proper motion in gamma-ray analyses generally requires evaluating the particle distribution resulting from continuous injection along the pulsar trajectory \cite{Tang-Piran-2019-posfrac-Geminga}, integrating it along the line of sight (LOS), and convolving the resulting projected distribution with the ICS kernel to obtain the gamma-ray surface brightness. This procedure can be computationally expensive, potentially limiting the inclusion of proper motion in extensive studies of TeV halos with current and future observations.\\

The aim of this work is to derive an analytical prescription for the
gamma-ray morphology of TeV halos around moving pulsars, avoiding
computationally expensive numerical integrations.
Our goal is to provide a practical means of constructing spatial
templates for gamma-ray observations, extending the analytical
approaches commonly adopted for stationary TeV halos
\citep[see, e.g.,][]{HAWC-2017-Geminga-Monogem, Rigoselli-2025-PoS-TeV-halos}.

We begin by re-deriving the three-dimensional lepton distribution around
a moving source, following \citet{Tang-Piran-2019-posfrac-Geminga}.
This solution involves a time integral over the particle injection
history. We then show that the integration along the line of sight
can be performed exactly, retaining the full geometrical dependence
on the source trajectory and observer position. The projected lepton
distribution is thus reduced to a single time integral, which can be
evaluated numerically for arbitrary injection spectra and energy-loss rates. The expression can be further simplified in the far-field limit, when the source distance is much larger than the characteristic size
of the halo.

Finally, adopting the far-field limit, we assume a constant injection
luminosity and a simplified treatment of radiative energy losses, following the prescription introduced by \citet{Atoyan-1995-long}
and commonly adopted in TeV-halo analyses
\citep{HAWC-2017-Geminga-Monogem}.
Under these assumptions, we show that the remaining time integral
can also be approximated analytically, yielding a fully analytical
expression for the projected lepton distribution.
The resulting solution naturally factorizes into an energy-dependent
spectral term, containing the injection spectrum and diffusion coefficient, and a dimensionless spatial term describing the
energy-dependent morphology and its dependence on the diffusion coefficient and pulsar velocity.

To facilitate their application, we provide a Python implementation
of the exact and approximate solutions, together with an illustrative
example, as supplementary material.

The paper is organized as follows. In Sect.~\ref{sec:leton-distr},
we derive the three-dimensional lepton distribution around a moving pulsar.
In Sect.~\ref{sec:gamma-morph-calc}, we perform the exact line-of-sight
integration, introduce the far-field approximation, and derive a fully
analytical expression for the projected lepton distribution under the
simplifying assumptions discussed above.
In Sect.~\ref{sec:validation}, we assess the accuracy of the approximations
and illustrate their application to the gamma-ray morphology of TeV halos,
using LHAASO J0248+6021, associated with PSR J0248+6021
\citep{LHAASO-2025-TeV-halo-PSRJ0248+6021}, as a reference case.
Finally, we summarize our results in Sect.~\ref{sec:conclusion}.

\section{Electron-positron distribution: continuous injection from a moving source with diffusion and losses}
\label{sec:leton-distr}

The distribution function for leptons instantaneously released at time $t_{\rm inj}$ from a point source located at position $\mathbf{x}_s$ and observed at time $t_{\rm obs}$ reads \citep{Atoyan-1995-long}
\begin{equation}
f_e(t_{\rm obs},\mathbf{x},E)
=
\frac{Q(E_{\tau})}{\pi^{3/2}r_d^3(E,E_{\tau})}
\frac{b(E_{\tau})}{b(E)}
\exp\left[
-\frac{|\mathbf{x}-\mathbf{x}_s|^2}{r_d^2(E,E_{\tau})}
\right],
\end{equation}
where $b(E)=-dE/dt$ is the positive energy-loss rate, $E_{\tau}$ is the initial energy of particles cooling down to energy $E$ over a time $\tau=t_{\rm obs}-t_{\rm inj}$, and $r_d$ is the diffusion-loss radius, defined by
\begin{equation*}
\tau = \int_E^{E_{\tau}}\frac{dE'}{b(E')}, \qquad \quad  r_d^2(E,E_{\tau}) = 4\int_E^{E_{\tau}}\frac{D(E')}{b(E')}\,dE'.
\end{equation*}

We model a pulsar of age $t_{\rm age}$, moving at a uniform velocity $\mathbf{v}$, as a sequence of burst-like sources injecting particles at times $t_i$ and positions $\mathbf{x}_{s,i} = \mathbf{x}_{\rm age} +  \mathbf{v}(t_i - t_{\rm age})$, where in the latter equality we assumed that the pulsar is located at position $\mathbf{x}_{s, \rm age}$ at present time.

The resulting distribution is the sum of this individual contributions,
\begin{equation}\label{eq:fe-sum-ti}
f_e(t_{\rm age},\mathbf{x},E)
=
\sum_i
\frac{Q(E_{\tau_i},t_i)}{\pi^{3/2}r_d^3(E,E_{\tau_i})}
\frac{b(E_{\tau_i})}{b(E)}
\exp\left[
-\frac{|\mathbf{x}-\mathbf{x}_{s,i}|^2}
{r_d^2(E,E_{\tau_i})}
\right],
\end{equation}
where $\tau_i=t_{\rm age}-t_i$. In addition, we assume that the source injects particles with a time-dependent spectrum $q(E,t)$, that can be normalized  to the injected particle luminosity  $L_e(t)$ according to $\int q(E,t)E\,dE=L_e(t)$. For each injection interval $\Delta t$, the corresponding burst-like injection spectrum is $Q(E_{\tau_i},t_i)=q(E_{\tau_i},t_i)\Delta t$.

Substituting this expression into Eq.~(\ref{eq:fe-sum-ti}) and taking the limit $\Delta t\to 0$, the sum becomes an integral over the injection time. We then change the integration variable from $t$ to $\tau=t_{\rm age}-t$ and obtain
\begin{align}\label{eq:fe-final}
f_e(t_{\rm age},\mathbf{x},E)
=
\int_0^{t_{\rm age}}d\tau\,
&\frac{q(E_{\tau},t_{\rm age}-\tau)}
{\pi^{3/2}r_d^3(E,E_{\tau})}
\frac{b(E_{\tau})}{b(E)}\\ \nonumber
& \times\exp\left[
-\frac{|\mathbf{x} - \mathbf{x}_{\rm age}+\mathbf{v}\tau|^2}
{r_d^2(E,E_{\tau})}
\right].
\end{align}

The energy loss rate due to synchrotron and inverse-Compton losses, with the addition of ionization, Coulomb and Bremsstrahlung losses, is calculated following \cite{Atoyan-1995-long}, while the Klein-Nishina effect in ICS is taken into account following 
\cite{Fang-2021-Klein-Nishina}. 

\section{Derivation of the gamma-ray morphology}
\label{sec:gamma-morph-calc}

The gamma-ray surface brightness is obtained by integrating the particle distribution of Eq.~(\ref{eq:fe-final}) over the radiative production rate and along the LOS:
\begin{equation}
\mathrm{SB}(t_{\rm age},\mathbf{x}_{\perp},E_{\gamma})
=
\int_0^{+\infty}ds\int dE\,
\frac{f_e(t_{\rm age},\mathbf{x}_{\rm los},E)}{4\pi}
\mathrm{PR}_{\gamma}(E_{\gamma},E,\mathbf{x}_{\rm los}),
\end{equation}
where $\mathbf{x}_{\rm los}=\mathbf{x}_{\rm los}(s,\mathbf{x}_{\perp})$ denotes a position along the LOS, $\mathbf{x}_{\perp}$ is the projected position on the plane of the sky at the source distance, and $\mathrm{PR}_{\gamma}$ is the differential $\gamma$-ray production rate.

For inverse-Compton emission, the target radiation fields, particularly the CMB, can be approximated as spatially uniform. Under this assumption, the spatial dependence of the emission is entirely determined by the particle distribution, and the surface brightness can be written as
\begin{equation}\label{eq:SB-Sigma}
\mathrm{SB}(t_{\rm age},\mathbf{x}_{\perp},E_{\gamma})
=
\int dE\,
\frac{\mathrm{PR}_{\gamma}(E_{\gamma},E)}{4\pi}
\Sigma(t_{\rm age},\mathbf{x}_{\perp},E),
\end{equation}
where $\Sigma (t_{\rm age},\mathbf{x}_{\perp},E)$ is the LOS-integrated particle distribution:
\begin{align}\label{eq:sigma-fe-los-integ-def}
\Sigma(t_{\rm age},\mathbf{x}_{\perp},E)
=
\int_0^{+\infty}ds
\int_0^{t_{\rm age}} & d\tau\,
\frac{A(\tau,E)}{\pi^{3/2}r_d^3(E,E_{\tau})} \\ \nonumber
& \times  \exp\left[
-\frac{|\mathbf{x}_{\rm los}(s,\mathbf{x}_{\perp}) -\mathbf{x}_{\rm age}+\mathbf{v}\tau|^2}
{r_d^2(E,E_{\tau})}
\right],
\end{align}
with
\begin{equation}
A(\tau,E)
=
q(E_{\tau},t_{\rm age}-\tau)
\frac{b(E_{\tau})}{b(E)}.
\end{equation}

\begin{figure}
\centering
\includegraphics[width=\hsize]{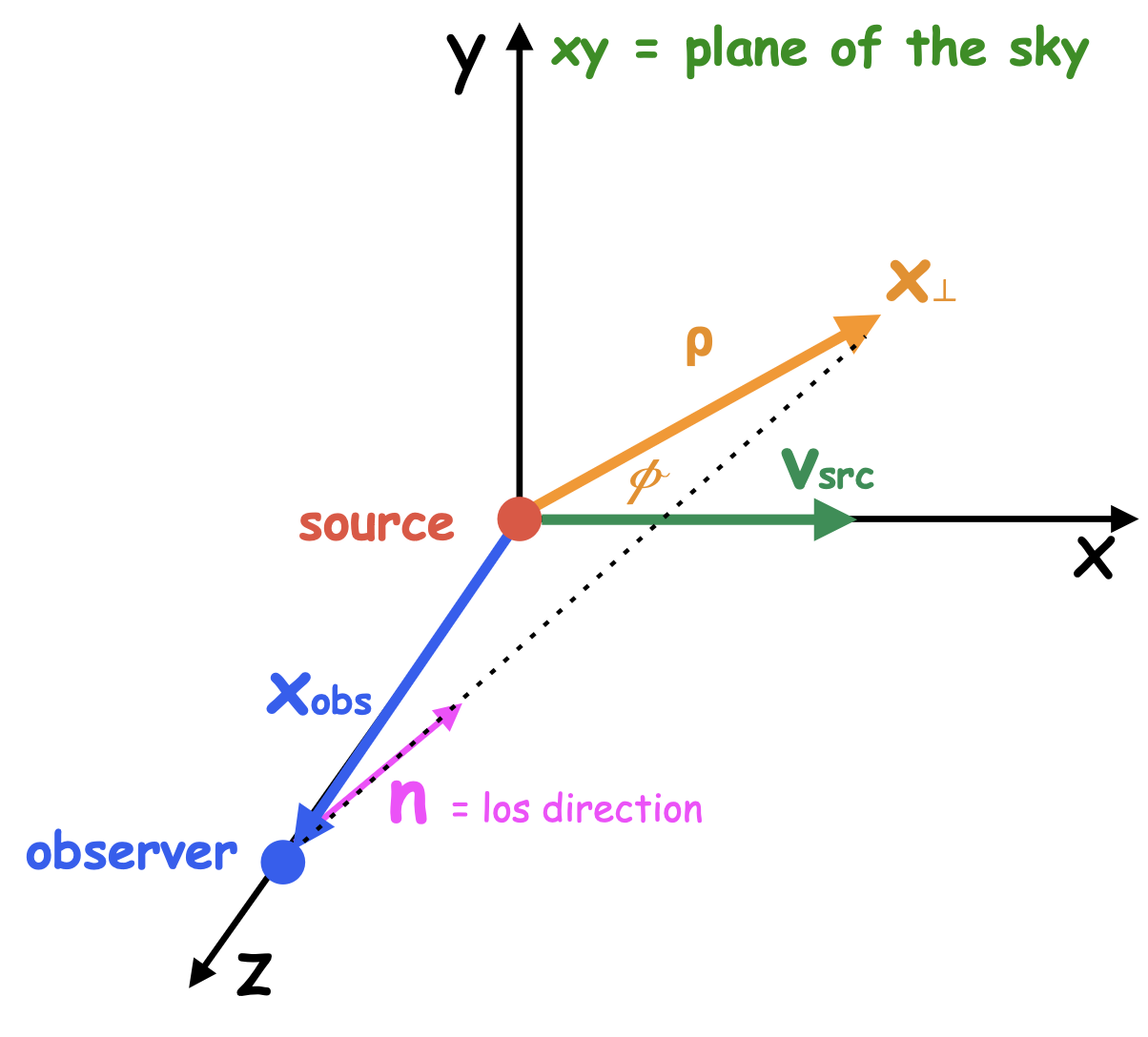}
\caption{Sketch of the adopted geometry. The current pulsar position is at the origin, its velocity is directed along the $x$-axis, and the observer is located at a distance $d$ along the $z$-axis.}
\label{fig:geom-sketch}
\end{figure}

To evaluate the LOS-integrated particle distribution, we consider the geometry illustrated in Fig.~\ref{fig:geom-sketch} and derive three expressions, whose detailed derivations are presented in Appendix~\ref{app:derivation}:

(i) \textbf{Exact LOS integration}: an exact expression for the LOS
integral, that retains the full dependence on the
pulsar and observer positions, the direction of pulsar motion, the injection history, and radiative energy losses.
The resulting projected distribution requires only a
one-dimensional numerical integration over the particle
injection history.

(ii) \textbf{Far-field approximation}: a simplified expression for the projected distribution, valid when the source distance is much larger than the characteristic halo extension, while retaining the full treatment of injection history and energy losses.

(iii) \textbf{Fully analytical approximation}: an approximate expression for the remaining time integral, valid in the far-field limit under the additional assumptions of constant injection luminosity and a simplified treatment of energy losses.

\subsection{Exact expression for the line-of-sight integrated particle distribution}

As shown in Appendix~\ref{app:derivation}, the integration along the LOS (variable $s$) in  Eq.~(\ref{eq:sigma-fe-los-integ-def}) is analytic and gives
\begin{align}
\Sigma(t_{\rm age},\mathbf{x}_{\perp},E)
=
\int_0^{t_{\rm age}}d\tau\,
&\frac{A(\tau,E)}{2\pi r_d^2(E,E_{\tau})}
\nonumber\\
&\times
\exp\left[-\frac{a^2-a_n^2}{r_d^2(E,E_{\tau})}\right]
\operatorname{erfc}\left(\frac{a_n}{r_d}\right),
\label{eq:Sigma-fe-los-integ-exact}
\end{align}
with
\begin{equation}
\mathbf{n} = \frac{\mathbf{x}_{\perp} - \mathbf{x}_{\rm obs}}{\vert \mathbf{x}_{\perp} - \mathbf{x}_{\rm obs} \vert}, \qquad 
\mathbf{a}=\mathbf{x}_{\rm obs} -\mathbf{x}_{\rm age} +\mathbf{v}\tau,
\qquad
a_n=\mathbf{n}\cdot\mathbf{a}.
\end{equation}
Here, $\mathbf{n}$ is the unit vector along the LOS, pointing from the observer towards the position $\mathbf{x}_{\perp}$ on the plane of the sky. 

This expression is exact and valid for a generic position of the observer $\mathbf{x}_{\rm obs}$ relative to the current position of the pulsar $\mathbf{x}_{\rm age}$. It can be combined with Eq.~(\ref{eq:SB-Sigma}) to compute the $\gamma$-ray surface brightness for a generic injection spectrum and time-dependent injection luminosity $L_e(t)$, using packages such as \texttt{NAIMA} \citep{naima} to evaluate the inverse-Compton emission.

\subsection{Far-field approximation for the line-of-sight integrated particle distribution}
\label{subsec:far-field}

Without loss of generality, and referring to the geometry illustrated in Fig.~\ref{fig:geom-sketch},
we place the observer along the $z$-axis, namely $\mathbf{x}_{\rm obs} = d\,\hat e_z$, and assume that the pulsar moves along the $x$-axis and is currently located at the origin, namely $\mathbf{x}_{\rm age} = \mathbf{0}$ and $\mathbf{v}\tau = v\tau\,\hat e_x$.
We also introduce polar coordinates to identify a position in the plane of the sky with respect to the current pulsar position
\begin{equation}
    \mathbf{x}_{\perp} = \rho\cos\phi\,\hat e_x +\rho\sin\phi\,\hat e_y.
\end{equation}

With these choices and assuming the far-field limit, $\rho\ll d$, $v\tau\ll d$ and $r_d \ll d$, one can show (see Appendix~\ref{app:derivation}) that the LOS-integrated particle distribution of Eq.~(\ref{eq:Sigma-fe-los-integ-exact})  simplifies to
\begin{align}
\Sigma(t_{\rm age},\rho,\phi,E)
=
\int_0^{t_{\rm age}} & d\tau\,
\frac{A(E, \tau)}{\pi r_d^2(E,E_\tau)}
\nonumber\\
&\times
\exp\left[
-\frac{\rho^2+(v\tau)^2+2\rho v\tau\cos\phi}
{r_d^2(E,E_\tau)}
\right].
\label{eq:Sigma-fe-far-field}
\end{align}

\subsection{Analytical approximation for constant injection luminosity and simplified energy losses}
\label{subsec:final-approx}

We now seek a fully analytical approximation to the LOS-integrated particle distribution. Following the treatment of radiative losses proposed by \citet{Atoyan-1995-long}, we assume that particles of energy $E$ contribute to the distribution over a maximum time
\begin{equation}
\tau_{\max}(E)
=
\min\left[t_{\rm age},\,\kappa\tau_{\rm loss}(E)\right],
\end{equation}
where $\kappa\simeq0.75$. Within this interval, we approximate $E_\tau\simeq E$ and $r_d^2(E,E_\tau)\simeq4D(E)\tau$. Note that the same assumption leads to the well-known approximation
of \citet{Atoyan-1995-long} for the electron distribution around a
stationary source with constant injection luminosity and isotropic
diffusion:
\begin{equation}
    f_e(t,r,E) =
    \frac{Q(E)}{4\pi D(E)r}
    \operatorname{erfc}\left(
        \frac{r}{\sqrt{4D(E)\tau_{\rm max}(E)}}
    \right).
\end{equation}
This expression has been widely adopted in analyses of TeV halos
\citep{HAWC-2017-Geminga-Monogem}.\\

Assuming a constant injection luminosity, one gets $A(E, \tau)\simeq q(E)$ and the far-field expression of Eq.~(\ref{eq:Sigma-fe-far-field}) reduces to
\begin{align}
\Sigma(t_{\rm age},\rho,\phi,E)
\simeq
\frac{q(E)}{4\pi D(E)}
&\exp\left[-\frac{\rho v\cos\phi}{2D(E)}\right]
\nonumber\\
&\times
\int_0^{\tau_{\max}(E)}
\frac{d\tau}{\tau}
\exp\left[
-\frac{\tau_D}{\tau}
-\frac{\tau}{\tau_{vD}}
\right],
\label{eq:Sigma-fe-los-integ-approx-nocool}
\end{align}
where we have introduced the characteristic timescales
\begin{equation}
\label{eq:timescale-expression}
\left\{
\begin{aligned}
&\tau_D(\rho,E) = \frac{\rho^2}{4D(E)},\\
&\tau_v(\rho) = \frac{\rho}{v},\\
&\tau_{vD}(E) = \frac{4D(E)}{v^2}.
\end{aligned}
\right.
\end{equation}
These represent, respectively, the diffusion time over a distance $\rho$, the time required for the pulsar to travel the same distance, and the characteristic timescale associated with the competition between diffusion and source motion.\\

As shown in Appendix~\ref{app:derivation}, the time integral in Eq.~(\ref{eq:Sigma-fe-los-integ-approx-nocool}) can be computed analytically in the cases of negligible proper motion $v\to0$ and of infinite $\tau_{\rm max}$, giving
\begin{equation}
\Sigma_{v=0}
=
\frac{q(E)}{4\pi D(E)}
E_1\left(\frac{\tau_D}{\tau_{\max}}\right).
\label{eq:Sigma-v0}
\end{equation}
and 
\begin{equation}
\Sigma_{\tau_{\rm max} \to \infty}
=
\frac{q(E)}{2\pi D(E)}
\exp\left[-\frac{\rho v\cos\phi}{2D}\right]
K_0\left(\frac{\rho v}{2D(E)}\right),
\label{eq:Sigma-infinite-time}
\end{equation}
where $E_1$ is the first-order exponential integral and $K_0$ is the modified Bessel function of the second kind.

Note that the limit $v \to 0$ recovers the projected particle distribution of a spherically symmetric TeV halo centered on a stationary source, providing an analytical parameterization that can also be used to model such systems.\\

For a finite $\tau_{\rm max}$ and a generic $v$, in Appendix~\ref{app:derivation} we show that the time integral can be approximated analytically, giving
\begin{align}
\Sigma_{\rm approx}(t_{\rm age},\rho,\phi,E)
\simeq
&\frac{q(E)}{4\pi D(E)}
\exp\left[-\frac{\rho v\cos\phi}{2D(E)}\right]
\nonumber\\
& \times K_0\left(\frac{\rho v}{2D}\right)
\left[
1+\operatorname{erf}
\left(
\frac{\ln(\chi_{\rm max}/\eta)}{w(\eta)}
\right)
\right],
\label{eq:Sigma-analytical-approx}
\end{align}
where 
\begin{equation}
\eta=\frac{\tau_v}{\tau_D},
\qquad
\chi_{\rm max}=\frac{\tau_{\max}}{\tau_D}, \qquad w(\eta)=
\frac{2e^{2/\eta}}{\sqrt{\pi}}
K_0\left(\frac{2}{\eta}\right).
\end{equation}

Note that the result of Eq.~(\ref{eq:Sigma-analytical-approx}) naturally splits into a spectral component $\mathcal{S}(E)$ and a dimensionless spatial factor $\mathcal{T}(\rho,\phi,E)$ that depends on energy  through the diffusion coefficient $D(E)$ and $\tau_{\rm max}(E)$:
\begin{equation}
\label{eq:Sigma-analytical-approx-split}
\left\{
\begin{aligned}
&\Sigma_{\rm approx}(t_{\rm age},\rho,\phi,E) = \mathcal{S}(E) \mathcal{T}(\rho,\phi,E),\\
&\mathcal{S}(E) = \frac{q(E)}{4\pi D(E)},\\
&\mathcal{T}(\rho,\phi,E) = \exp\left[-\frac{\rho v\cos\phi}{2D}\right] K_0\left(\frac{\rho v}{2D}\right)
\left[
1+\operatorname{erf}
\left(
\frac{\ln(\chi_{\rm max}/\eta)}{w(\eta)}
\right)
\right].
\end{aligned}
\right.
\end{equation}

\subsection{Equilibrium lepton spectrum and total gamma-ray flux}
\label{sec:equilibrium-spectrum}

The spatially-integrated lepton spectrum is obtained by integrating
the projected lepton density over the entire plane of the sky. This corresponds to the equilibrium lepton spectrum, as shaped by injection and energy losses. 
As shown in Appendix~\ref{app:equilibrium-spectrum}, this yields
\begin{equation}
N_{\rm eq}(E)
=
\int_0^{t_{\rm age}}d\tau\,
q(E_\tau,\tau)\frac{b(E_\tau)}{b(E)}.
\label{eq:equilibrium-spectrum-main}
\end{equation}
The result is independent of the diffusion coefficient and
pulsar velocity, since spatial transport redistributes the
particles without changing their total number at a given energy.
In the simplified treatment, this reduces to
\begin{equation}
N_{\rm eq}(E)\simeq q(E)\tau_{\max}(E).
\label{eq:equilibrium-spectrum_approx-main}
\end{equation}

The total gamma-ray flux produced by inverse-Compton scattering at distance $d$,
\begin{equation}
\frac{d\Phi_\gamma}{dE_\gamma}
=
\frac{1}{4\pi d^2}
\int dE\,
N_{\rm eq}(E)\,
{\rm PR}_{\gamma, \rm IC}(E_\gamma,E),
\label{eq:total-gamma-flux}
\end{equation}
can therefore be calculated directly from the equilibrium lepton
spectrum $N_{\rm eq}(E)$, for instance using \texttt{NAIMA}.

\section{Validation of the approximations}
\label{sec:validation}

In this section, we assess the accuracy of the analytical approximations introduced in the previous sections by comparing their predictions with the exact calculation of the projected electron distribution reported of Eq.~(\ref{eq:Sigma-fe-los-integ-exact}). 

As a representative case, we consider the extended multi-TeV gamma-ray source
LHAASO J0248+6021 \citep{LHAASO-2025-TeV-halo-PSRJ0248+6021},
associated with PSR J0248+6021 (spin-down power $\dot{E} \approx 2\times10^{35}\,\mathrm{erg\,s^{-1}}$, age $t_{\rm age} \approx 60\,{\rm kyr}$, distance $d\sim 2\,{\rm kpc}$).
This system is particularly interesting for investigating the effects of
pulsar proper motion, owing to the relatively high pulsar transverse velocity, $v_{\rm psr} \sim 500\,\mathrm{km\,s^{-1}}$, and the diffusion coefficient
inferred from the LHAASO observations,
$D(E) \approx 2\times10^{28}(E/160\,\mathrm{TeV})^{0.5}
\,\mathrm{cm^2\,s^{-1}}$.
The combination of substantial pulsar velocity and relatively slow
particle diffusion makes the effects of proper motion potentially relevant
for the spatial distribution of the gamma-ray emission.

Energy losses include synchrotron emission in a magnetic field of $3\,\mu\mathrm{G}$ and inverse-Compton scattering on the CMB and on the interstellar radiation field components considered in \cite{Fang-2021-Klein-Nishina}.

The corresponding LHAASO measurements are included as an observational reference for the characteristic spectral and spatial properties of the emission. We emphasize that the adopted parameters are not obtained through a fit to the data, and that the purpose of this exercise is to test the approximations rather than to construct a detailed model of the source.\\

\begin{figure}[ht!]
    \centering
    \includegraphics[width=\hsize]{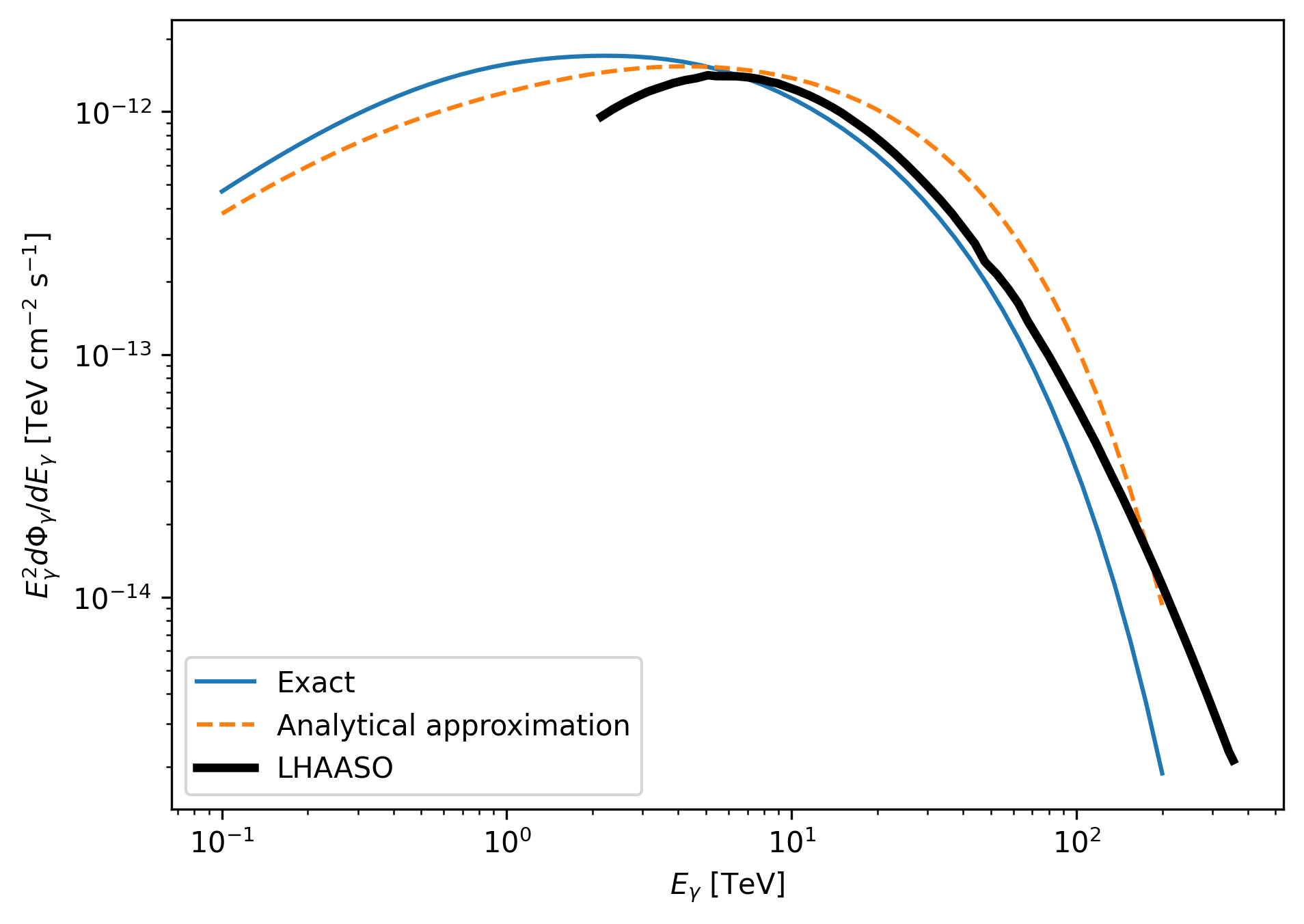}
    \caption{
    Inverse-Compton gamma-ray spectral energy distributions obtained
    from the exact calculation (solid blue line) and the analytical
    approximation (dashed orange line).
    The best-fit spectral model for a log-parabola function,  reported by LHAASO for J0248+6021 is shown as a reference curve (black solid line).
    The differences between the predicted spectra arise from the
    approximate treatment of electron energy losses, which becomes
    less accurate near the spectral cutoff.
    }
    \label{fig:gamma-total-sed}
\end{figure}

We assume continuous injection of electron--positron pairs with a power-law
spectrum and an exponential cutoff, adopting a constant injection luminosity
of $L_e = 2\times10^{34}\,\mathrm{erg\,s^{-1}}$, corresponding to approximately
$10\%$ of the current pulsar spin-down power, a spectral index of $1.2$,
and a cutoff energy of $E_{\rm cut}=50\,\mathrm{TeV}$.
These parameters are chosen to approximately reproduce the overall spectral energy distribution (SED) reported by LHAASO, following the discussion of Sect.~\ref{sec:equilibrium-spectrum} and without performing a detailed fit.\\
Figure~\ref{fig:gamma-total-sed} compares the gamma-ray SEDs obtained
from the exact and approximate calculations, together with the
best-fit spectral model reported by LHAASO for a log-parabola function, shown for reference.
The two calculations are in good overall agreement, although small
differences emerge around the spectral cutoff.
These differences arise from the simplified treatment of energy losses
in the approximate calculation, 
which does not fully reproduce the
equilibrium electron spectrum in this energy range.\\

\begin{figure*}[ht!]
    \centering
    \includegraphics[width=\textwidth]{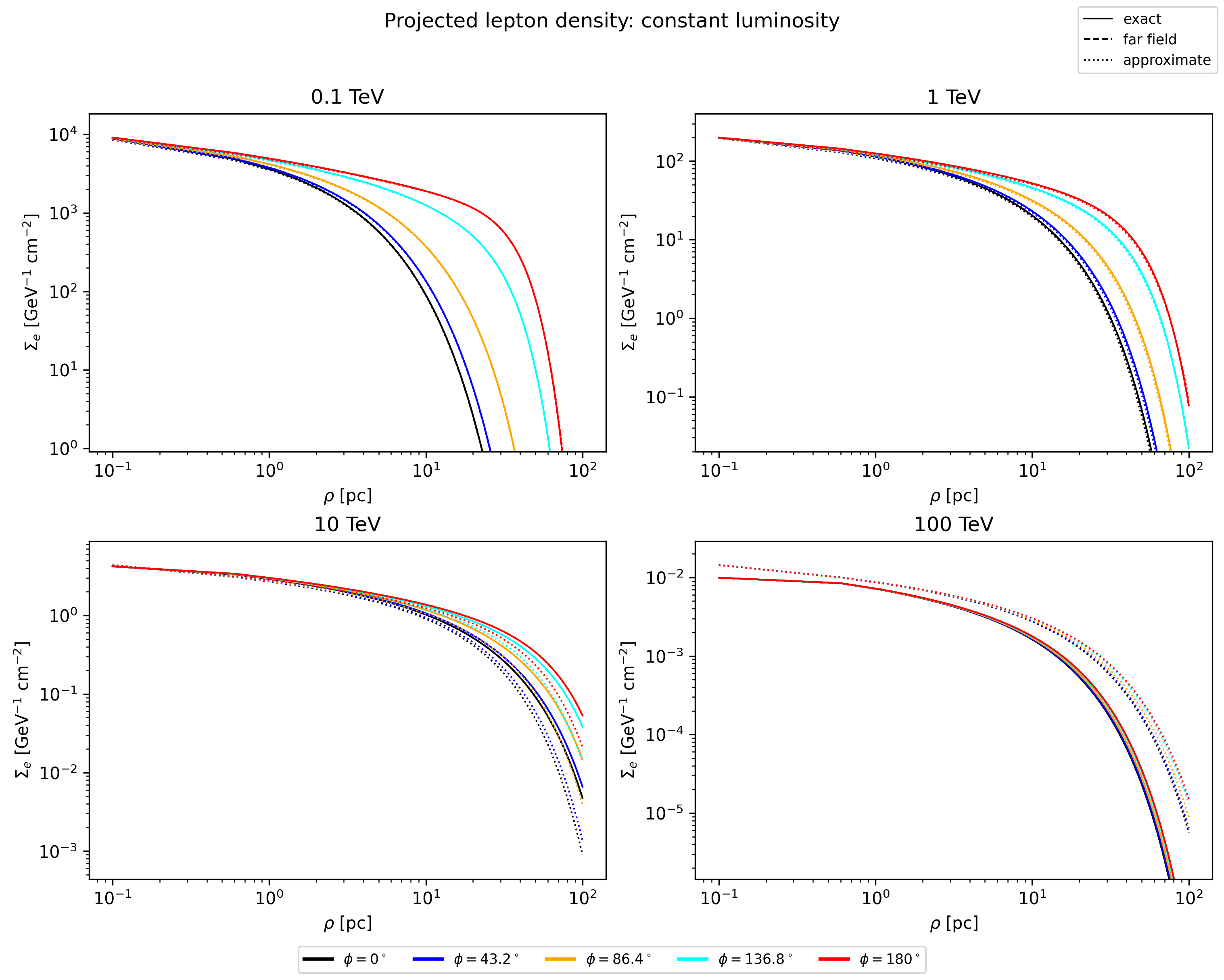}
    \caption{
    Radial profiles of the projected electron density
    $\Sigma_e$ at different electron energies, obtained using
    the exact calculation (solid lines), the far-field
    approximation (dashed lines), and the analytical approximation
    (dotted lines).
    Different colors correspond to different azimuthal angles
    relative to the direction of pulsar motion (as marked).
    The comparison illustrates the accuracy of the approximations
    in reproducing both the normalization and spatial dependence
    of the electron distribution, including the asymmetry
    induced by the pulsar proper motion.
    }
    \label{fig:lepton-radial-profiles}
\end{figure*}

In Fig.~\ref{fig:lepton-radial-profiles}, we compare the projected electron
distributions obtained with the exact calculation, the far-field approximation,
and the analytical approximation. The comparison is performed at representative
electron energies between $0.1$ and $100\,\mathrm{TeV}$, and along different
azimuthal directions relative to the pulsar proper motion.

As discussed in Sect.~\ref{app:derivation}, the accuracy of the analytical
approximation of the time integral depends on the parameter
\begin{equation}
    \eta = \frac{4D(E)}{\rho v}
    \approx 25\,\frac{E_{\rm TeV}^{0.5}}{\rho_{\rm pc}\,v_{500}},
\end{equation}
Good agreement is found for $\eta < 10$, with particularly high accuracy
for $\eta < 0.1$. Although deviations increase for larger values of
$\eta$, they remain below 10\% up to $\eta \sim 100$, as shown in Sect.~\ref{app:derivation}. Larger discrepancies occur only in regions
where the integral is already close to zero, and therefore have little
impact on the resulting particle distribution.
For the reference values considered here, $\eta$ can significantly exceed
unity, particularly at small projected distances and high electron energies.
However, even in these regimes, the approximation generally remains
within the range of good accuracy identified above, with larger deviations
limited to regions where the projected particle density is strongly
suppressed.

As expected, the projected electron distribution is asymmetric, extending
to larger distances behind the moving pulsar. The three calculations show
remarkably good agreement at energies below $\sim 1\,\mathrm{TeV}$ across
the entire considered spatial range. At higher energies, deviations become
more pronounced, particularly at projected distances exceeding
$\sim 10\,\mathrm{pc}$.

In general, the analytical approximation becomes less accurate at large
distances and at energies approaching the cutoff of the injected spectrum.
As already illustrated by the comparison of the total gamma-ray SEDs,
these discrepancies are primarily associated with the simplified treatment
of electron energy losses, which becomes less reliable near the spectral
cutoff.\\

\begin{figure*}[ht!]
    \centering
    \includegraphics[width=\textwidth]{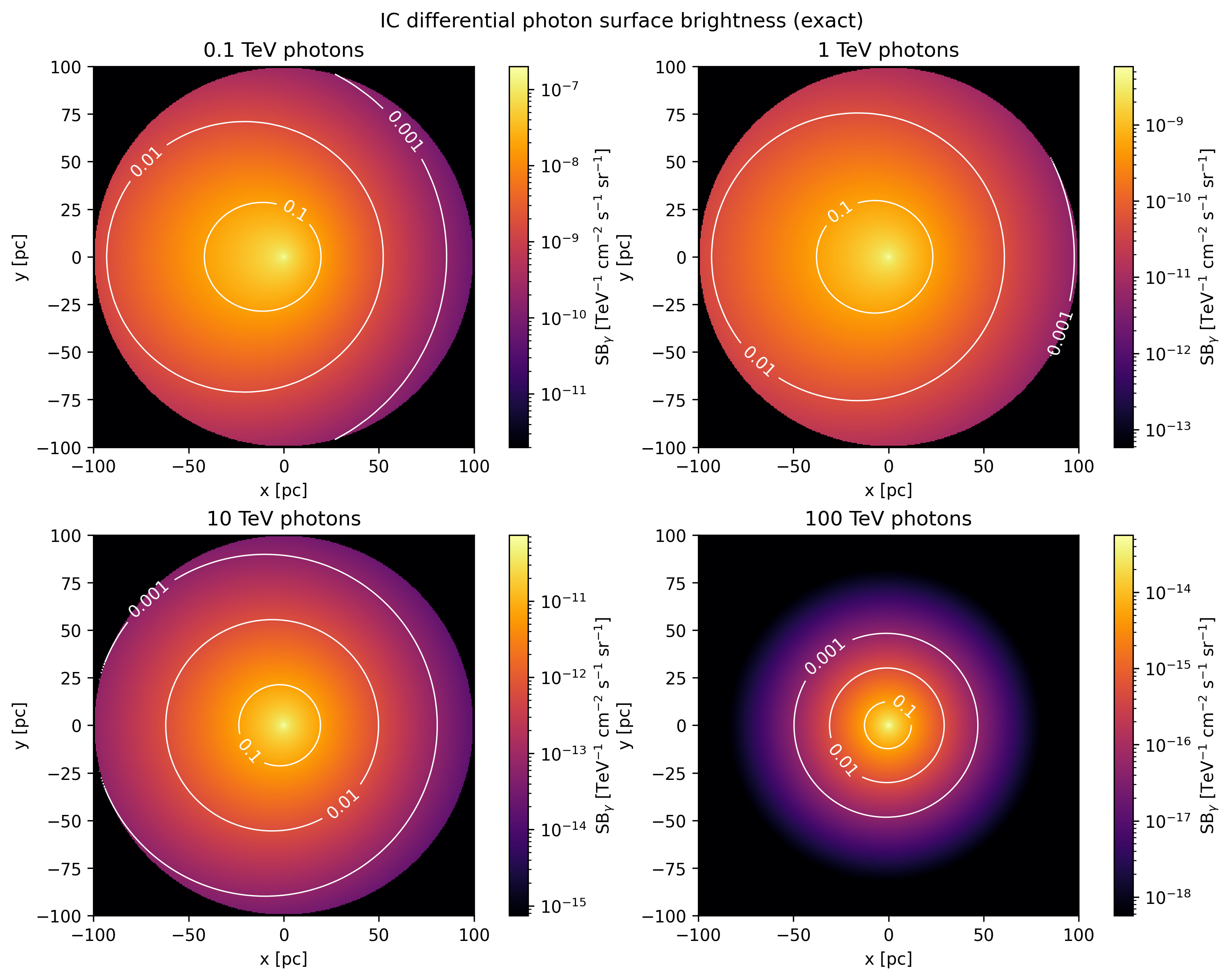}
    \caption{
    Inverse-Compton gamma-ray surface-brightness maps at representative
    photon energies of $0.1$, $1$, $10$, and $100\,\mathrm{TeV}$,
    calculated using the exact projected electron distribution.
    The color scale represents the differential gamma-ray surface
    brightness, while the contours indicate levels of
    $0.1\%$, $1\%$, and $10\%$ of the maximum brightness in each panel.
    The maps illustrate the energy-dependent asymmetry induced by
    pulsar proper motion, which becomes more pronounced at lower
    photon energies.
    }
    \label{fig:gamma-sb-map-contour-exact}
\end{figure*}

In Fig.~\ref{fig:gamma-sb-map-contour-exact}, we show the inverse-Compton
gamma-ray surface-brightness maps at representative photon energies between
$0.1$ and $100\,\mathrm{TeV}$, calculated using the exact LOS-integrated electron distribution. The emission is computed using the same
radiation fields adopted in the calculation of electron energy losses.
The maps clearly shows an asymmetric morphology, with the emission extending
preferentially behind the moving pulsar. This asymmetry is more pronounced
at lower photon energies, where longer electron cooling times and smaller
diffusion coefficients allow particles injected at earlier stages of the
pulsar evolution to retain a stronger spatial imprint of its past positions.\\

\begin{figure}[ht!]
    \centering
    \includegraphics[width=\hsize]{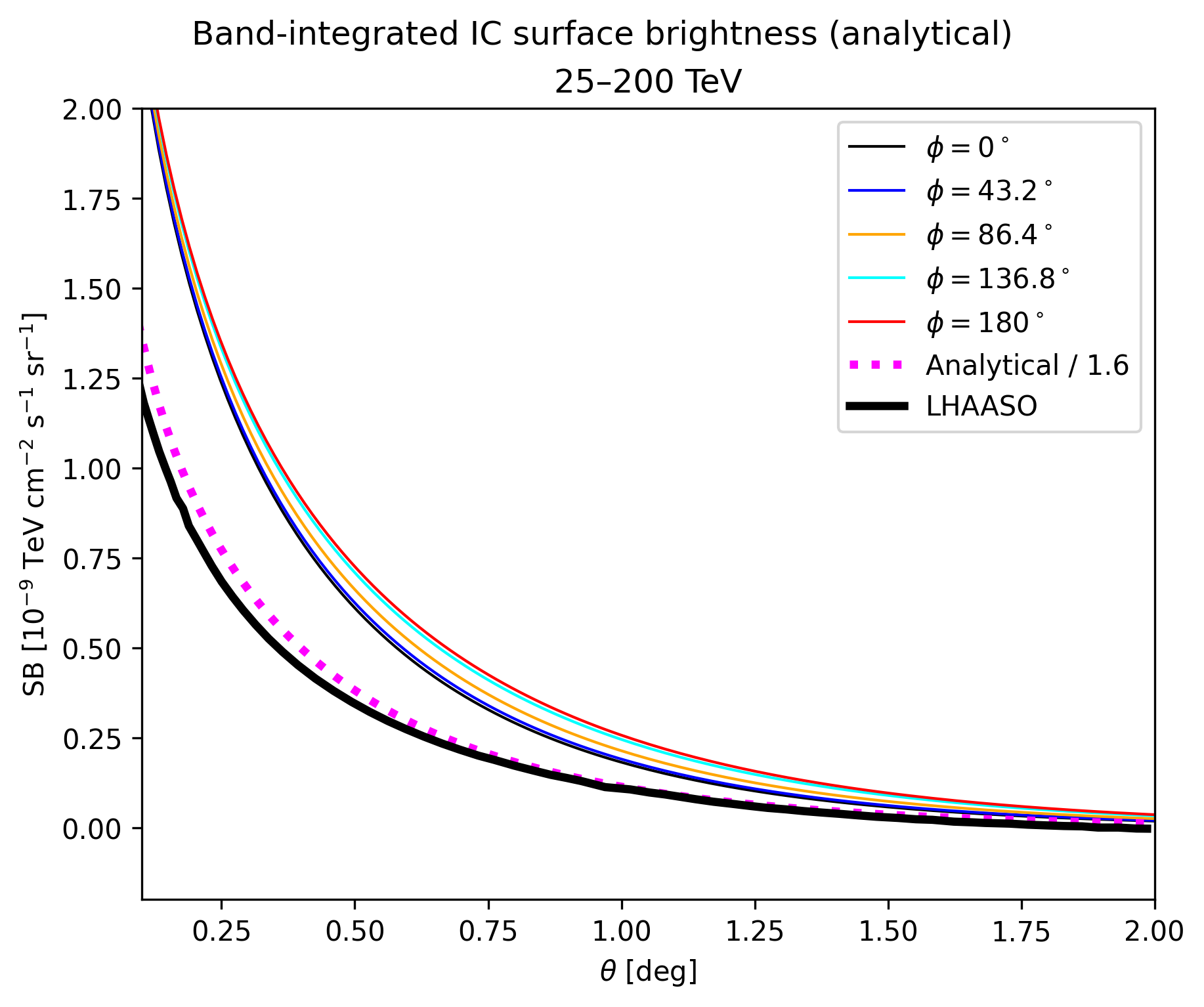}
    \includegraphics[width=\hsize]{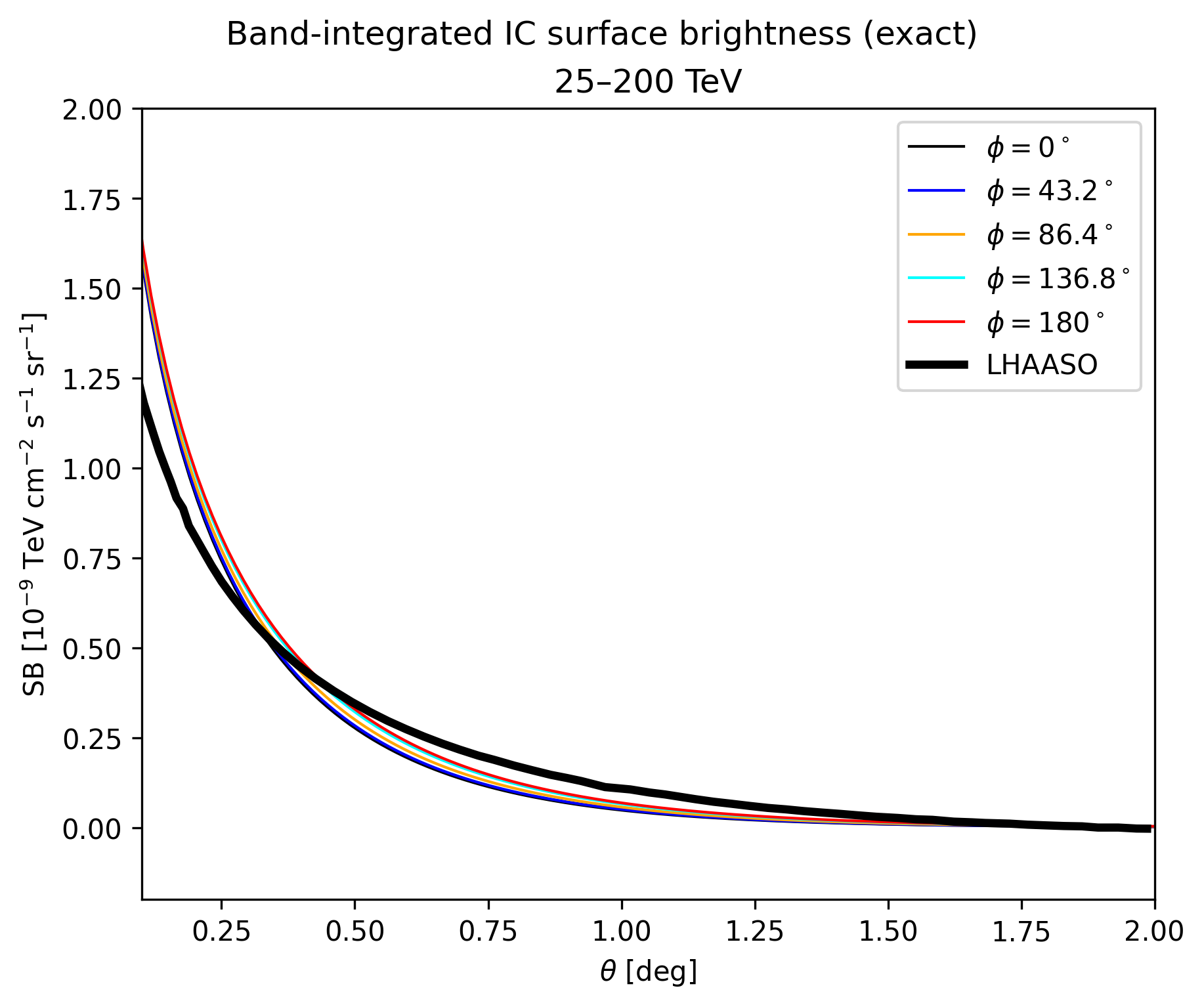}
    \caption{
    Radial profiles of the inverse-Compton gamma-ray surface brightness
    integrated over the photon-energy range $25$--$200\,\mathrm{TeV}$,
    calculated using the analytical approximation (upper panel)
    and the exact projected electron distribution (lower panel).
    Different colors correspond to different azimuthal directions
    relative to the pulsar proper motion.
    The best-fit radial profile reported by LHAASO for J0248+6021,
    obtained using their diffusion model, is shown for reference.
    In the upper panel, the analytical prediction rescaled by a factor
    of $1/1.6$ is also shown, illustrating that the discrepancy
    with the exact calculation is predominantly associated with
    the overall normalization.
    The model predictions are not convolved with the instrumental PSF,
    and no fit to the LHAASO observations has been performed.
    }
    \label{fig:gamma-integrated-sb-profiles}
\end{figure}

To further investigate the spatial distribution of the gamma-ray emission,
we consider the surface brightness integrated over the photon-energy range
$25$--$200\,\mathrm{TeV}$.
In Fig.~\ref{fig:gamma-integrated-sb-profiles}, we compare the resulting
radial profiles with the best-fit one-dimensional gamma-ray flux distribution
reported by LHAASO \citep{LHAASO-2025-TeV-halo-PSRJ0248+6021}
for $E_{\gamma}> 25~\rm TeV$, obtained using their diffusion model.
Also in this case, the LHAASO profile is shown for reference only,
and no attempt is made to fit the observational data.

For the illustrative parameters adopted here, the exact calculation
provides a reasonable description of the normalization and angular
dependence of the LHAASO reference profile.
The analytical approximation exhibits larger deviations, consistent
with the discrepancies already identified in the total gamma-ray SED
at energies approaching the spectral cutoff.
Interestingly, these differences primarily affect the overall normalization,
while the radial dependence is reasonably well reproduced.
This is illustrated by the additional curve obtained by dividing the
integrated surface brightness predicted by the analytical approximation
by a factor of $\sim 1.6$, which brings it into closer agreement with
the exact calculation.
This suggests that, in the energy range considered, the main limitation
of the analytical approximation lies in the simplified treatment of
electron energy losses rather than in the description of the spatial
morphology. Such a discrepancy could introduce systematic uncertainties
in the inferred particle injection parameters, particularly in the
lepton injection efficiency relative to the pulsar spin-down power.

Finally, we note that the predicted profiles have not been convolved
with the instrumental point-spread function (PSF).
Consequently, the agreement with the LHAASO reference profile should
be interpreted as illustrative, and not as a quantitative assessment
of the source morphology.

\section{Conclusions}
\label{sec:conclusion}

We have derived analytical expressions for the projected distribution of
electrons and positrons continuously injected by a moving pulsar and
propagating through the surrounding medium under the effects of
energy-dependent diffusion and radiative losses.

In particular, we have
shown that the integration of the particle density along the line of sight
can be performed exactly, without approximations on the observer and source positions, and on the particle injection history and radiative energy losses. This result reduces the
calculation of the projected particle distribution to a single time
integral, eliminating the need for an additional, computationally
expensive spatial integration. It therefore provides an efficient and
accurate way to calculate the gamma-ray surface brightness of extended
sources associated with moving pulsars, while retaining the full effects
of their proper motion and injection history.

We have further simplified the exact solution by introducing the
far-field approximation, valid when the source distance is much larger
than the characteristic size of the halo and the pulsar displacement.
Under the additional assumptions of constant injection luminosity
and a simplified treatment of radiative energy losses, we have
derived a fully analytical expression for the projected lepton
distribution, eliminating the remaining time integral.
The resulting expression factorizes into an energy-dependent spectral
term, containing the injection spectrum and diffusion coefficient,
and a dimensionless spatial term describing the energy-dependent
morphology and its dependence on the diffusion coefficient and
pulsar velocity.
This provides a practical analytical parameterization of the
asymmetric emission around moving pulsars.
In the limit of vanishing pulsar velocity, the expression also
provides an analytical parameterization of the projected lepton
distribution of spherically symmetric TeV halos
around stationary pulsars.

By comparing the exact and approximate solutions, we have assessed
the accuracy of the approximations and identified their main
limitations, particularly at large distances from the source, where
the emission is generally fainter, and at high energies, especially
near the spectral cutoff, where the simplified treatment of
radiative losses becomes less reliable.
We have also shown that the latter primarily affects the inferred
lepton injection spectrum, particularly its normalization, when
the model is compared with gamma-ray observations.

We have illustrated the results through the calculation of the
inverse-Compton gamma-ray emission from a representative TeV-halo source,
using parameters inspired by  LHAASO J0248+6021, associated to PSR J0248+6021. The comparison confirms that the exact line-of-sight integration provides a practical means of
calculating energy-dependent gamma-ray morphologies, while the analytical
approximations reproduce their main spatial features with reasonable
accuracy. These results are particularly relevant for interpreting
extended gamma-ray sources associated with pulsars, where proper motion
can introduce significant departures from spherical symmetry.

To facilitate the application of these results, we provide a Python
implementation of the exact and approximate solutions, together with
an illustrative example, as supplementary material.

\bibliographystyle{aa} % style aa.bst
\bibliography{biblio.bib} % your references Yourfile.bib

\begin{appendix}
\section{Analytical derivation of the projected lepton distribution}
\label{app:derivation}

The starting point is the LOS-integrated particle distribution given in Eq.~(\ref{eq:sigma-fe-los-integ-def}). In what follows, we derive: \textit{(i)} an exact expression for the LOS integral; \textit{(ii)} a far-field approximation to the LOS integral, valid when the source distance is much larger than the characteristic halo extension; and \textit{(iii)} an analytical approximation for the remaining time integral, valid in the far-field limit under the assumptions of constant injection luminosity and a simplified treatment of energy losses.

\subsection{Exact expression for the line-of-sight integrated particle distribution}

To perform the LOS integration in Eq.~(\ref{eq:sigma-fe-los-integ-def}), we consider the geometry illustrated in Fig.~\ref{fig:geom-sketch}. 

\noindent
A position along a LOS can be expressed in terms of the observer position $\mathbf{x}_{\rm obs}$ and the LOS unit vector $\mathbf{n}$ as
\begin{equation}
\mathbf{x}_{\rm los}(s,\mathbf{x}_{\perp})
=
\mathbf{x}_{\rm obs}+s\mathbf{n},
\end{equation}
where $\mathbf{n}$ can be expressed in terms of  $\mathbf{x}_{\perp}$ and we introduce the vector $\mathbf{a}$ :
\begin{equation}
\mathbf{n} = \frac{\mathbf{x}_{\perp} - \mathbf{x}_{\rm obs}}{\vert \mathbf{x}_{\perp} - \mathbf{x}_{\rm obs} \vert}, \qquad 
\mathbf{a}=\mathbf{x}_{\rm obs} -\mathbf{x}_{\rm age} +\mathbf{v}\tau,
\qquad
a_n=\mathbf{n}\cdot\mathbf{a}. 
\end{equation}
The spatial dependence of the of the exponent can be written as
\begin{equation}
|\mathbf{x}_{\rm los} -\mathbf{x}_{\rm age} +\mathbf{v}\tau|^2
=
(s+a_n)^2+a^2-a_n^2,
\end{equation}
which allows for the  LOS integration to be performed analytically using
\begin{equation}
\int_0^{+\infty}ds\,
e^{-(s+a_n)^2/r_d^2}
=
\frac{\sqrt{\pi}}{2}r_d\,
\operatorname{erfc}\left(\frac{a_n}{r_d}\right).
\end{equation}
We thus obtain
\begin{align}
\Sigma(t_{\rm age},\mathbf{x}_{\perp},E)
=
\int_0^{t_{\rm age}}d\tau\,
&\frac{A(\tau,E)}{2\pi r_d^2(E,E_{\tau})}
\nonumber\\
&\times
\exp\left[-\frac{a^2-a_n^2}{r_d^2(E,E_{\tau})}\right]
\operatorname{erfc}\left(\frac{a_n}{r_d}\right).
\end{align}

\subsection{Far-field approximation for the line-of-sight integrated particle distribution}

Without any loss of generality, we introduce the geometry illustrated in Fig.~\ref{fig:geom-sketch} and described in Sect~\ref{subsec:far-field}. A position on the plane of the sky and the associated LOS direction can be expressed in polar coordinates as
\begin{align}
& \mathbf{x}_{\perp} = \rho\cos\phi\,\hat e_x +\rho\sin\phi\,\hat e_y,  \\ \nonumber
& \mathbf{n} =
\frac{\rho\cos\phi\,\hat e_x+\rho\sin\phi\,\hat e_y - d\hat e_z}{\sqrt{\rho^2+d^2}}.
\end{align}
Recalling the definitions of $\mathbf{a}$ and $a_n$ from the general case, we obtain
\begin{equation}
a^2 = d^2+(v\tau)^2, \qquad \quad
a_n = \frac{\rho v\tau\cos\phi-d^2}
{\sqrt{\rho^2+d^2}}.
\end{equation}
In the far-field limit, $\rho\ll d$ and $v\tau\ll d$ and the relevant geometrical quantities reduce to
\begin{equation}
a^2-a_n^2
\simeq \rho^2+(v\tau)^2+2\rho v\tau\cos\phi.
\end{equation}
Furthermore, since the diffusion-loss radius is much smaller than the source distance, $r_d\ll d$, and $a_n \approx -d$ the complementary error function in Eq.~(\ref{eq:Sigma-fe-los-integ-exact}) satisfies
\begin{equation}
\operatorname{erfc}\left(\frac{a_n}{r_d}\right)
\simeq \operatorname{erfc}\left(-\frac{d}{r_d}\right)
\simeq 2.
\end{equation}

\noindent
The LOS-integrated particle distribution therefore simplifies to
\begin{align}
\Sigma(t_{\rm age},\rho,\phi,E)
=
\int_0^{t_{\rm age}} & d\tau\,
\frac{A(E, \tau)}{\pi r_d^2(E,E_\tau)}
\nonumber\\
&\times
\exp\left[
-\frac{\rho^2+(v\tau)^2+2\rho v\tau\cos\phi}
{r_d^2(E,E_\tau)}
\right].
\end{align}

\subsection{Analytical approximation for constant injection luminosity and simplified energy losses}

Following Sect.~\ref{subsec:final-approx}, we start from Eq.~(\ref{eq:Sigma-fe-los-integ-approx-nocool})
and consider the timescales defined in Eq.~(\ref{eq:timescale-expression}).

\subsubsection{Negligible proper motion.}
In the limit $v\to0$, we have $\tau_{vD}\to\infty$, and the time integral reduces to 
\begin{equation}
\int_0^{\tau_{\max}(E)}
\frac{d\tau}{\tau}
\exp\left[-\frac{\tau_D(\rho, E)}{\tau}\right]
\end{equation}
which can be solved analytically with a suitable change of variable  and  by using the definition of the first-order exponential integral (defined for $x>0$)
\begin{equation}
\int_{x}^{\infty}\frac{d u}{u}e^{-u}
=E_1(x).
\end{equation}
Finally we obtain:
\begin{equation}
\Sigma_{v=0}
=
\frac{q(E)}{4\pi D(E)}
E_1\left(\frac{\tau_D}{\tau_{\max}}\right).
\end{equation}

\subsubsection{Infinite integration time.}
In the limit $\tau_{\max}\to\infty$, the integral can instead be evaluated in terms of the modified Bessel function of the second kind, $K_0$,
\begin{equation}
 \int_0^{\infty}\frac{d\tau}{\tau}
\exp\left[-\frac{a}{\tau}-\frac{\tau}{b}\right]
=2K_0\left(2\sqrt{\frac{a}{b}}\right),  
\end{equation}
yielding
\begin{equation}
\Sigma_{\tau_{\rm max} \to\infty}
=
\frac{q(E)}{2\pi D(E)}
\exp\left[-\frac{\rho v\cos\phi}{2D(E)}\right]
K_0\left(\frac{\rho v}{2D(E)}\right).
\end{equation}
We will use this limiting solution to construct an analytical approximation for finite $\tau_{\max}$.

\subsubsection{Approximation for finite integration time.}
To approximate the time integral in Eq.~(\ref{eq:Sigma-fe-los-integ-approx-nocool}), we introduce the dimensionless quantities
\begin{equation}
\chi=\frac{\tau}{\tau_D},
\qquad
\eta=\frac{\tau_v}{\tau_D},
\qquad
\chi_{\rm max}=\frac{\tau_{\max}}{\tau_D}.
\end{equation}
The integral then becomes
\begin{equation}
I(\eta,\chi_{\rm max})
=
\int_0^{\chi_{\rm max}}\frac{d \chi}{\chi}
\exp\left[-\frac{1}{\chi}-\frac{\chi}{\eta^2}\right],
\label{eq:I-eta-X}
\end{equation}
with
\begin{equation}
I_\infty(\eta)
=
2K_0\left(\frac{2}{\eta}\right),
\end{equation}
as found in the infinite-integration-time limit.

We define the normalized cumulative integral
\begin{equation}
G(\eta, \chi_{\rm max})
=
\frac{I(\eta, \chi_{\rm max})}{I_\infty(\eta)},
\end{equation}
which increases from zero to unity as $\chi_{\rm max}$ increases. Introducing the variable
\begin{equation}
z=\ln\left(\frac{\chi}{\eta}\right),
\end{equation}
the cumulative normalized integral can be rewritten as
\begin{equation}
G(\eta, \chi_{\rm max})
=
\frac{1}{I_\infty(\eta)}
\int_{-\infty}^{\ln(\chi_{\rm max}/\eta)}
dz\,\exp\left[-\frac{2}{\eta}\cosh z\right].
\label{eq:G-cosh}
\end{equation}

The integrand is symmetric around $z=0$ and becomes increasingly peaked at small $\eta$. In this regime, the quadratic expansion $\cosh z\simeq 1+z^2/2$ provides a good Gaussian approximation. At larger $\eta$, however, the integrand becomes broader, and this approximation significantly overestimates the integral, since the quadratic expansion underestimates $\cosh z$ away from $z=0$. To obtain a more accurate approximation, we instead introduce a Gaussian with an effective width $w(\eta)$, determined by requiring that the integral of $I(\eta , \chi_{\rm max})$ over the entire domain reproduces the exact result $I_\infty(\eta)$. 

This gives
\begin{equation}
w(\eta)
=
\frac{2e^{2/\eta}}{\sqrt{\pi}}
K_0\left(\frac{2}{\eta}\right)
\end{equation}
and
\begin{equation}
G(\eta, \chi_{\rm max})
\simeq
\frac{1}{2}
\left[
1+\operatorname{erf}
\left(
\frac{\ln(\chi_{\rm max}/\eta)}{w(\eta)}
\right)
\right],
\label{eq:G-approx}
\end{equation}
where we used
\begin{equation}
\int_{-\infty}^t dx\,e^{-x^2/\eta^2}
= \frac{\sqrt{\pi} \eta}{2} \left[ 1 + \operatorname{erf}\!\left(\frac{t}{\eta}\right)\right]
\end{equation}

\begin{figure*}[ht!]
    \centering
     %\resizebox{12cm}{12cm}
    {\includegraphics {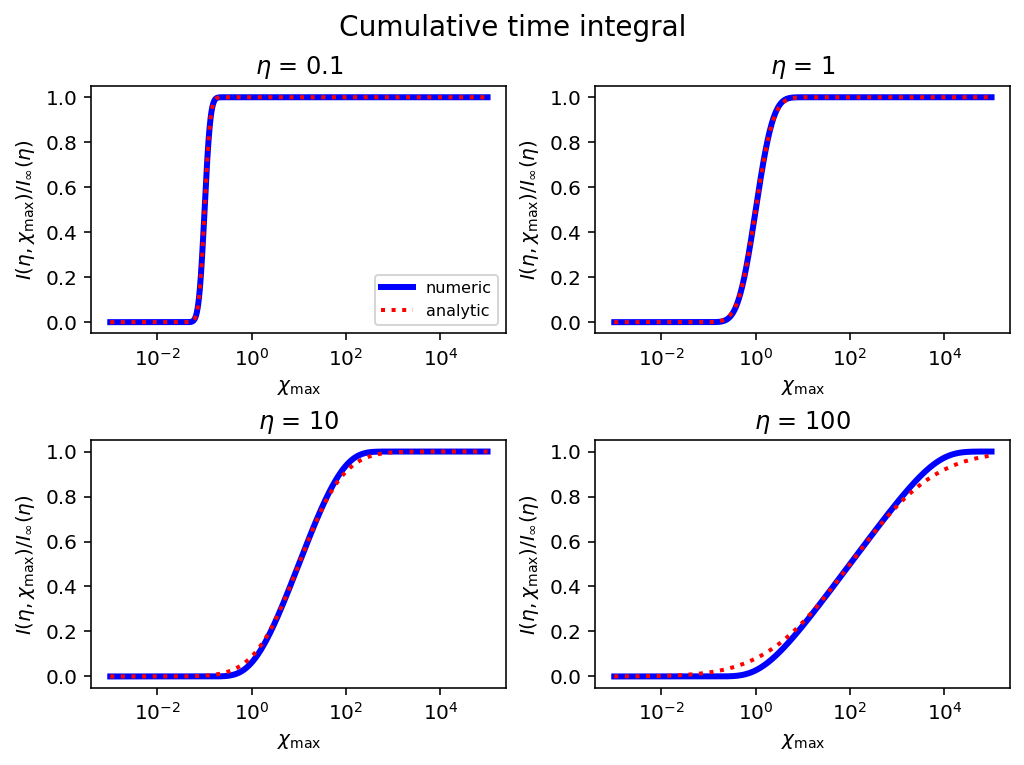}}
     \caption{Comparison between the numerical evaluation (blue solid lines) and the analytical approximation (red dotted lines) of the cumulative integral $I(\eta,\chi_{\rm max})$ of Eq.~(\ref{eq:I-eta-X}), for different values of $\eta$. The integral is normalized to its asymptotic value, $I_\infty(\eta)=2K_0(2/\eta)$, where $K_0$ is the modified Bessel function of the second kind. The comparison illustrates the accuracy of the analytical approximation as a function of the upper integration limit $\chi_{\rm max}$ and the of the parameter $\eta$.}
      \label{fig:integral-comp}
\end{figure*}

Therefore the integral can be approximated as 
\begin{equation}
I(\eta, \chi_{\rm max})
\simeq
K_0\left(\frac{2}{\eta}\right)
\left[
1+\operatorname{erf}
\left(
\frac{\ln(\chi_{\rm max}/\eta)}{w(\eta)}
\right)
\right].
\label{eq:I-approx}
\end{equation}
The accuracy of this approximation is illustrated in Fig.~\ref{fig:integral-comp}, where the cumulative integral is compared with its numerical evaluation for different values of $\eta$. The agreement is excellent for $\eta\lesssim 1$, where the integrand
is strongly localized around its maximum and is well approximated by a Gaussian in $\ln(\chi)$. As $\eta$ increases, the integrand becomes broader and progressively departs from a Gaussian shape,
leading to increasing differences between the analytical approximation and the numerical result. Nevertheless, the approximation remains
reasonably accurate up to $\eta\sim 10$, while more significant deviations appear at larger values. Even for $\eta=100$, however, the discrepancy remains below $\sim 10\%$ over the range of
$\chi_{\rm max}$ where the integral is appreciably different from zero. By construction, the approximation reproduces the exact asymptotic value $I_\infty(\eta)=2K_0(2/\eta)$ for all $\eta$; the discrepancies therefore concern the dependence of the cumulative integral on the upper integration limit $\chi_{\rm max}$. 

Substituting this approximation into Eq.~(\ref{eq:Sigma-fe-los-integ-approx-nocool}), we finally obtain
\begin{align}
\Sigma(t_{\rm age},\rho,\phi,E)
\simeq
\frac{q(E)}{4\pi D(E)}
&\exp\left[-\frac{\rho v\cos\phi}{2D(E)}\right]
\nonumber\\
&\times K_0\left(\frac{2}{\eta}\right)
\left[
1+\operatorname{erf}
\left(
\frac{\ln(\chi_{\rm max}/\eta)}{w(\eta)}
\right)
\right].
\end{align}

\section{Spatially integrated lepton spectrum}
\label{app:equilibrium-spectrum}

The spatially integrated lepton spectrum (equilibrium electron spectrum) is obtained by integrating
the projected lepton density over the plane of the sky. 

Starting from the far-field solution in
Eq.~\eqref{eq:Sigma-fe-far-field}, we obtain
\begin{align}
N_{\rm eq}(E)
&=
\int_0^{2\pi}d\phi
\int_0^\infty \rho\,d\rho\,
\Sigma(t_{\rm age},\rho,\phi,E)
\nonumber\\
&=
\int_0^{t_{\rm age}}d\tau\,
\frac{2A(E,\tau)}{r_d^2}
\exp\left[-\frac{(v\tau)^2}{r_d^2}\right]
\nonumber\\
&\quad\times
\int_0^\infty \rho\,d\rho\,
\exp\left[-\frac{\rho^2}{r_d^2}\right]
I_0\left(\frac{2\rho v\tau}{r_d^2}\right),
\end{align}
where the angular integration has been performed explicitly.

The remaining radial integral can be evaluated using the identity
\begin{equation}
\int_0^\infty dx\,x\,e^{-\alpha x^2}I_0(\beta x)
=
\frac{1}{2\alpha}
\exp\left(\frac{\beta^2}{4\alpha}\right),
\end{equation}
which gives
\begin{equation}
\int_0^\infty \rho\,d\rho\,
\exp\left[-\frac{\rho^2}{r_d^2}\right]
I_0\left(\frac{2\rho v\tau}{r_d^2}\right)
=
\frac{r_d^2}{2}
\exp\left[\frac{(v\tau)^2}{r_d^2}\right].
\end{equation}
Substituting this result into the previous expression yields
\begin{equation}
N_{\rm eq}(E)
=
\int_0^{t_{\rm age}}d\tau\,A(E,\tau).
\end{equation}

The dependence on the diffusion coefficient and pulsar velocity therefore disappears upon spatial integration. Physically,
diffusion and pulsar proper motion redistribute the particles
in space, while radiative losses modify their energies.
Since spatial transport conserves particle number, the same
result applies to the exact solution in
Eq.~\eqref{eq:Sigma-fe-los-integ-exact}, provided that the
integration encompasses the entire particle distribution.

Recalling the definition of $A(E,\tau)$, the equilibrium spectrum
is thus given by
\begin{equation}
N_{\rm eq}(E)
=
\int_0^{t_{\rm age}}d\tau\,
q(E_\tau,\tau)\frac{b(E_\tau)}{b(E)},
\end{equation}
where the integration is restricted to propagation times corresponding to physically allowed injection energies.

In the simplified treatment of energy losses and assuming a time-independent injection, this reduces to
\begin{equation}
N_{\rm eq}(E)\simeq q(E)\,\tau_{\max}(E),
\end{equation}
where $\tau_{\max}(E)$ is the maximum effective propagation time at energy $E$.

\end{appendix}

\end{document}